# Autonomous adaptive spectral analysis with an LLM agent for self-driving laboratories

Yuki K. Wakabayashi* and Yui Ogawa

Basic Research Laboratories, NTT, Inc., Atsugi, Kanagawa 243-0198, Japan

*Correspondence: yuuki.wakabayashi@ntt.com

## Abstract

Self-driving laboratories accelerate materials exploration through autonomous closed-loop integration of synthesis, characterization, and next-experiment selection. However, characterization does not always directly provide the target property; measured spectra or images often require automated analysis to extract quantitative descriptors based on physically motivated assumptions and predefined criteria. Such procedures can fail for measurements outside these criteria, requiring a decision on whether a reliable value can be recovered or no numerical value should be reported. Here, we implemented an autonomous adaptive spectral-analysis agent using a large language model (LLM) to supervise a deterministic numerical-analysis routine. For each spectrum, the agent executes the predefined analysis script and assesses the result using the measured spectrum and fitting diagnostics. Valid results are accepted unchanged, recoverable failures are reanalyzed through bounded adjustments of predefined fitting parameters, and NaN (not a number) is returned when no scientifically defensible value can be obtained. We evaluated the workflow on optical transmittance spectra from 203 epitaxial $\beta$-$Ga_2O_3$ films grown during a previously reported self-driving thin-film growth campaign. The predefined optical analysis script returned Urbach energy, an optical-quality metric, for 183 spectra and failed for 20. The agent retained those 183 values and evaluated 20 exception cases, recovering 16 through bounded parameter adjustments and returning NaN for 4 when the spectral evidence did not support a reliable Urbach tail. The audit records captured spectral features, parameter changes, and rationale for each decision. These results demonstrate autonomous scientific exception handling by an LLM agent while leaving quantitative property extraction to deterministic numerical analysis.

## 1. Introduction

Materials synthesis and processing often involve large multidimensional parameter spaces in which experiments are costly and exhaustive exploration is impractical.[1] Self-driving laboratories (SDLs) address this challenge by autonomously closing the loop among materials synthesis, characterization, data analysis, and algorithmic selection of subsequent experiments, using artificial intelligence and machine learning (AI/ML).[2–5] By iteratively directing experiments toward informative or promising regions of the parameter space, SDLs can accelerate the exploration and optimization of complex materials systems.[6–11] Such closed-loop approaches have been demonstrated across solution-based synthesis,[12] powder-based inorganic synthesis,[13] and thin-film growth,[14–19] enabling the discovery of new materials and synthesis routes as well as improvements in functional performance across semiconductors, battery materials,[20] and quantum materials.[21]

An essential component of this loop is the conversion of characterization data into quantitative values used for experimental decision making. Instruments often provide diffraction patterns, spectra, images, or electrical data rather than the target property itself.[22–27] These data are therefore processed using predefined analysis code or automated procedures to extract quantities such as peak positions,[28,29] band gaps, surface roughness, or carrier mobility. Such procedures encode physically motivated assumptions, fitting choices, and acceptance criteria. Because these procedures are fixed in advance, unexpected measurements can cause the analysis to fail and return no value. Some such failures may be recovered by modest adjustments of analysis parameters, whereas in other cases the measurement may not support a reliable quantitative estimate.

Even in an SDL, such exceptions often require human intervention. A researcher must then decide whether to adjust the analysis or leave the property value unreported. This intervention interrupts the autonomous experimental loop and can become a bottleneck in high-throughput operation.[18,30–33] An autonomous analysis layer should therefore distinguish between recoverable analysis failures and measurements that do not support a reliable property estimate, while recording any modifications and the rationale for them. Agents that combine LLMs with executable analysis tools provide a possible route to adaptive scientific decision making. Recent studies have begun to use such agents as an interface between scientific reasoning and executable workflows in chemistry and materials science.[34–38] LLMs have also been investigated for explaining synthesizability in inorganic crystals.[39] A practical question remains, however, when a fixed analysis procedure fails: should the procedure be adapted to obtain a scientifically defensible value, or should the system abstain from reporting one?

Here, we develop an LLM agent for autonomous adaptive spectral analysis coupled to a deterministic optical-analysis script for extracting the Urbach energy ($E_U$), an optical-quality metric characterizing the sub-band-gap absorption tail,[18] from optical transmittance spectra. For each spectrum, the agent evaluates the measured spectrum and analysis outputs, accepts the original analysis when appropriate, makes bounded adjustments to predefined analysis parameters when a recoverable failure is identified, and abstains when the measurement does not support a reliable estimate. Accordingly, all quantitative property values are calculated by the deterministic analysis script, while the agent provides the adaptive decision layer and records any parameter changes and the scientific rationale for its decisions. We evaluate this approach using optical transmittance spectra from 203 epitaxial films of $\beta$-$Ga_2O_3$, an ultrawide-bandgap semiconductor,[40] grown during a previously reported self-driving sputter-epitaxy campaign in which $E_U$ served as an objective for growth optimization.[18] The fixed analysis script returned finite $E_U$ values for 183 spectra and failed for 20. The agent retained the 183 values returned by the baseline analysis and evaluated the 20 exception cases, recovering 16 through bounded parameter adjustments while returning NaN (not a number) for the remaining 4 spectra when no scientifically defensible Urbach tail could be identified.

## 2. Methods

The dataset used for evaluation consisted of optical transmittance spectra from 203 epitaxial $\beta$-$Ga_2O_3$ films grown during the self-driving RF-sputtering campaign reported in Ref. **18**. The $\beta$-$Ga_2O_3$ films were grown on c-plane $Al_2O_3$ substrates under varying growth conditions, with substrate temperature, RF power, $O_2$ flow rate, and Ar flow rate as the controllable growth parameters. Each spectrum was stored as a CSV file and covered an energy range of approximately $h\nu$ = 2.07–6.53 eV, where $h$ is Planck's constant and $\nu$ is the photon frequency. No additional film growth or optical measurements were performed for the present study. To emulate the conditions of a new SDL campaign, in which spectra are acquired and analyzed sequentially, each archived spectrum was presented to the agent individually in a fresh session without information from other samples.

For each optical-transmission spectrum $T(h\nu)$, the agent first ran the original Python analysis script used in the SDL campaign to extract the $E_U$, using its default fitting parameters. The absorption coefficient $\alpha(h\nu)$ was estimated from the measured transmittance using the Beer–Lambert approximation,[41]

$$\alpha(h\nu) = -\frac{1}{d} ln[T(h\nu) + \varepsilon], \tag{1}$$

where $d$ is the nominal film thickness of 55 nm and $\varepsilon$ is a small constant to avoid singularities at $T \rightarrow 0$. The optical bandgap $E_g$ was then estimated using a direct-transition Tauc relation,[42–44]

$$(\alpha h\nu)^2 \propto h\nu - E_g, \tag{2}$$

by linear fitting in the appropriate region and taking the intercept with the $h\nu$ axis. Then, $E_U$ was extracted from the exponential absorption Urbach tail[45–47] just below the estimated $E_g$ by fitting

$$\ln[\alpha(h\nu)] = \frac{1}{E_U} h\nu + C, \tag{3}$$

where $C$ is a constant. In the default analysis, the first candidate Urbach-fitting window was defined as $E_g$ - 0.11 to $E_g$ - 0.03 eV. A candidate fit was accepted when it contained at least eight data points, yielded a positive fitted slope, and satisfied $R^2 \geq 0.995$. If these criteria were not met, progressively wider fitting windows were tested by shifting the lower-energy boundary to lower energies while keeping the high-energy boundary fixed at $E_g$ - 0.03 eV, up to a maximum window width of 0.40 eV, corresponding to lower-energy boundaries from $E_g$ - 0.11 to $E_g$ - 0.43 eV. The first interval satisfying these criteria was returned as the $E_U$ estimate.

The adaptive decision layer was implemented as a ChatGPT Workspace Agent using GPT-5.6 Sol, with the reasoning level set to very high. Figure 1 summarizes the decision workflow used for each spectrum, including acceptance of the original result, bounded parameter adjustment when a recoverable analysis failure was identified, and abstention when no scientifically defensible $E_U$ could be obtained. The decision protocol was defined in the agent instructions, and the original Python analysis script was provided as the executable numerical-analysis program. The complete agent instructions are provided in Supplementary Note S1. The agent was allowed to execute the analysis script, inspect the measured spectrum and fitting diagnostics, make temporary adjustments to predefined fitting parameters when justified, and record all modifications and decision rationales. Each spectrum was analyzed in an independent fresh session, and no parameter modification or decision was carried over between samples. The agent instructions prohibited web search and the use of external information. The agent was not given access to the previous SDL publication or to previously obtained analysis results, including archived $E_U$ values.

For each spectrum, the agent assigned one of three predefined decision labels: ACCEPT, REPAIR_ACCEPT, or NaN. ACCEPT retained the $E_U$ value obtained by running the original analysis script without modification when the result was supported by the spectrum and fit-quality metrics. When the original analysis failed, the agent assessed whether a scientifically reliable $E_U$ could be obtained through a small adjustment of the existing fitting parameters. If so, the agent modified the relevant parameter, reran the analysis, and reassessed the result. If a reliable $E_U$ value was obtained, the outcome was classified as REPAIR_ACCEPT. If the agent determined that no scientifically reliable $E_U$ could be obtained within the permitted repair process, NaN was assigned.

These adaptive repairs were deliberately constrained. The agent was instructed to modify the smallest number of fitting parameters necessary while leaving the raw spectrum and the definition of $E_U$ unchanged. Adjustments were limited to predefined parameters, including the fitting-window position and fit-quality threshold, and the fitting region was required to remain associated with the absorption edge rather than being moved elsewhere simply to obtain an acceptable fit. No more than three repair attempts were permitted for each spectrum, with each attempt consisting of a parameter adjustment followed by reanalysis. The agent could also examine nearby fitting windows to assess the local stability of a candidate $E_U$. These stability checks, together with the spectral shape and fit diagnostics, informed whether a repair was accepted or no scientifically defensible $E_U$ could be obtained. The complete repair rules and decision criteria are given in Supplementary Note S1. For each spectrum, the audit record documented the input spectrum and analysis script, analysis status, final $E_g$ and $E_U$ values when available, decision label, repair count, parameter changes, fit diagnostics, fitted interval, and the rationale for the final decision.

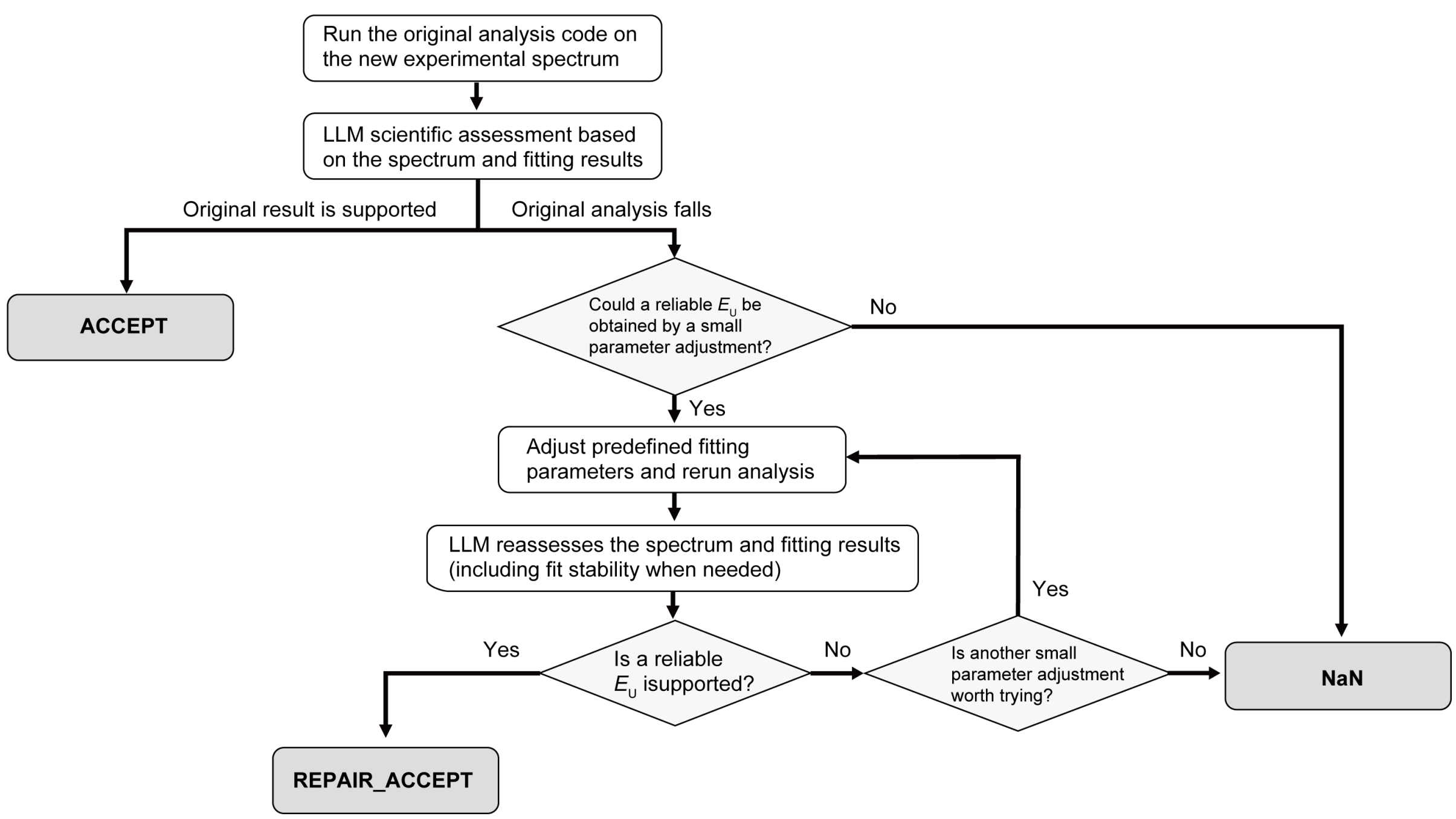

**Figure 1. Workflow of the autonomous adaptive spectral-analysis agent.** Each spectrum is first analyzed using the original analysis procedure, after which the LLM agent evaluates the measured spectrum and fitting results. Supported results are accepted without modification (ACCEPT). When the original analysis fails, the agent may make a bounded adjustment to predefined fitting parameters and reassess the resulting fit. A scientifically supported repaired result is classified as REPAIR_ACCEPT; otherwise, the agent may attempt another permitted adjustment or return NaN when no scientifically defensible $E_U$ can be obtained.

## 3. Results

The original analysis procedure returned an $E_U$ value for 183 of the 203 spectra and failed to identify an acceptable Urbach-fitting window for the remaining 20 spectra. After evaluation by the LLM agent, all 183 results returned by the original analysis procedure were retained without modification (ACCEPT), whereas 16 of the 20 failed cases were recovered through bounded parameter adjustment (REPAIR_ACCEPT) and 4 were assigned NaN (Fig. 2). The workflow therefore returned finite $E_U$ values for 199 of 203 spectra (98.0%) while assigning an explicit outcome to every spectrum. Thus, the adaptive decision process recovered quantitative $E_U$ values for 80% of the 20 exception cases, while avoiding unsupported numerical estimates for the remaining 20% when no scientifically defensible estimate could be obtained.

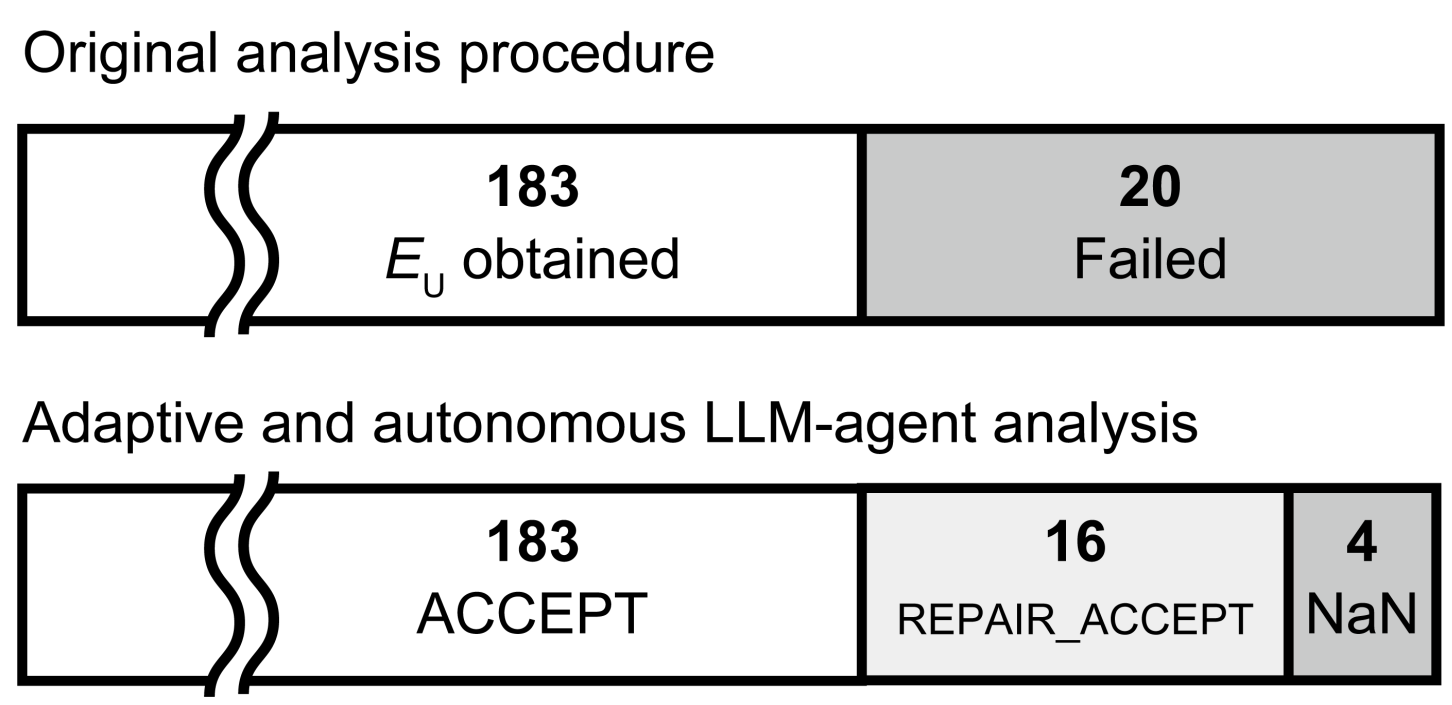

Figure 2. **Outcomes for the 20 spectra not resolved by the original analysis procedure.** The LLM agent recovered 16 cases through bounded parameter adjustment and assigned NaN to the remaining 4 cases.

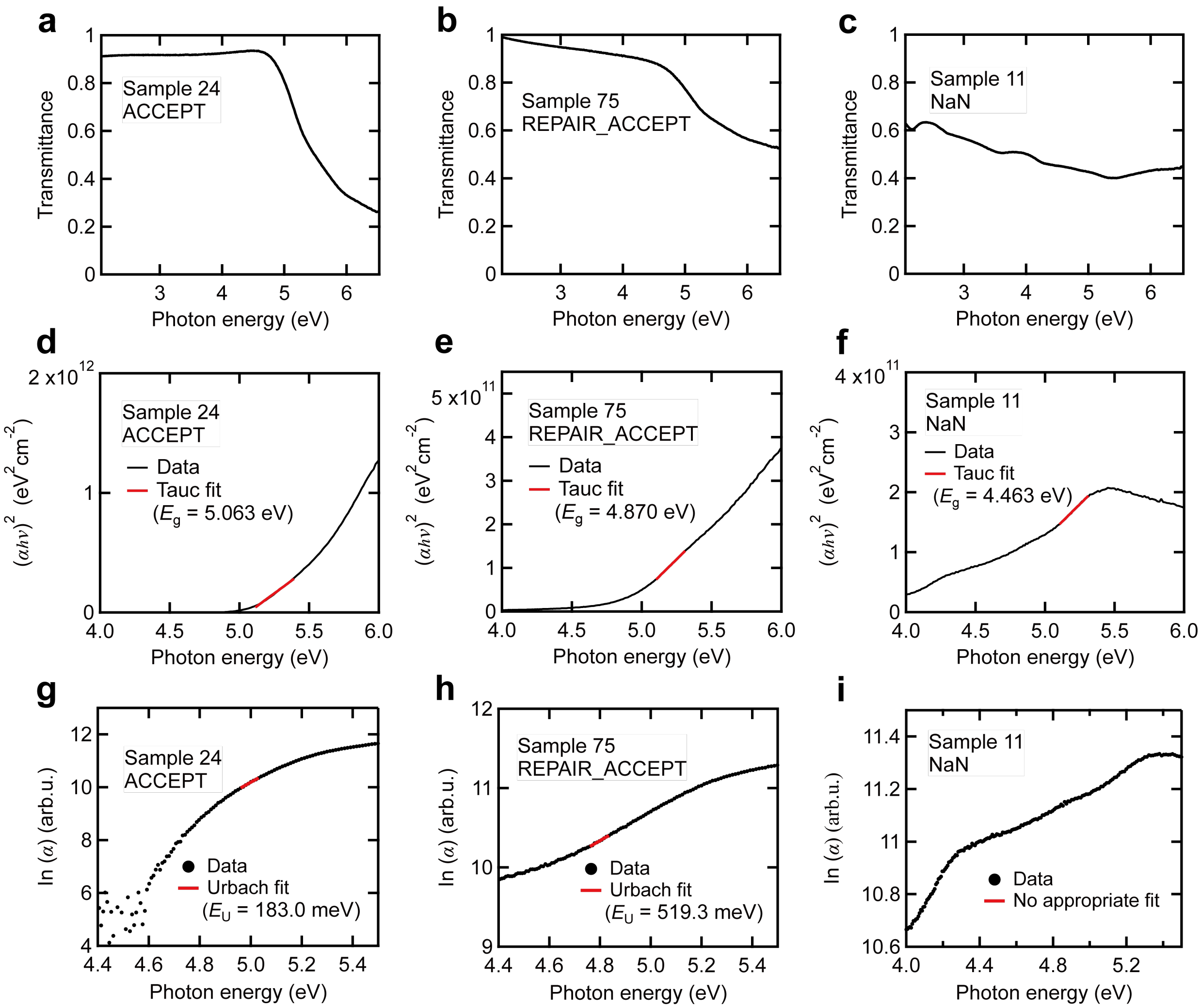

**Figure 3. Representative ACCEPT, REPAIR_ACCEPT, and NaN cases.** Results are shown for sample 24 (ACCEPT; left column), sample 75 (REPAIR_ACCEPT; center column), and sample 11 (NaN; right column). (a–c) Measured transmittance spectra. (d–f) Tauc plots used to estimate $E_g$, with linear fits shown in red. (g–i) $\ln[\alpha(h\nu)]$ spectra used for Urbach analysis, with the accepted fitting regions shown in red where applicable.

Figure 3 shows representative ACCEPT, REPAIR_ACCEPT, and NaN decisions. Sample 24 illustrates an ACCEPT decision. The transmittance spectrum shows a well-defined absorption edge (Fig. 3a), and the Tauc analysis gives $E_g$ = 5.063 eV (Fig. 3d). In the fitting window immediately below $E_g$, $\ln[\alpha(h\nu)]$ follows the linear dependence expected for an Urbach tail, yielding $E_U$ = 183.0 meV with $R^2$ = 0.997 (Fig. 3g). The agent's audit rationale explicitly noted that the fitted region was located immediately below the same absorption edge used to determine $E_g$, showed sufficient optical contrast and Urbach-like behavior, and was stable when the low- and high-energy boundaries of the Urbach-fitting window were each shifted by up to approximately ±0.02 eV, yielding $E_U$ values of 172.6–192.7 meV (Supplementary Table S1). The agent therefore retained the original result without modification. The other ACCEPT cases showed similar spectral and fitting characteristics and were also retained without modification.

Sample 75 illustrates a REPAIR_ACCEPT decision. The transmittance spectrum shows a well-defined absorption edge (Fig. 3b), and the Tauc analysis gives $E_g$ = 4.870 eV (Fig. 3e). However, the original analysis procedure failed because none of the candidate Urbach-fitting windows satisfied the predefined $R^2 \geq 0.995$ criterion. The agent therefore made a small adjustment to the fitting-window position, changing the high-energy boundary from $E_g$ - 0.03 to $E_g$ - 0.01 eV. The rerun yielded $E_U$ = 519.3 meV with $R^2$ = 0.99718 (Fig. 3h). The agent's audit rationale noted that the fitted region was located immediately below the absorption edge, showed monotonic Urbach-like behavior, and that the change in transmittance across the fitted region was clearly larger than the local measurement noise. The extracted $E_U$ was also stable against small variations in the Urbach-fitting parameters, with the high-energy boundary and fitting-window width varied by up to approximately 0.02 eV, giving values of 519.3–558.0 meV (Supplementary Table S1). The agent therefore classified the result as REPAIR_ACCEPT after one repair attempt.

Across the 16 REPAIR_ACCEPT cases (Supplementary Table S1), nine were recovered by relaxing only the predefined $R^2$ threshold from 0.995 to 0.989–0.994, two by shifting the high-energy boundary while retaining the original $R^2$ criterion, and one by adjusting the fitting-window width. The remaining four required a bounded combination of a boundary shift and $R^2$ relaxation. All successful repairs involved only predefined fitting parameters; none altered the measured data or introduced a new fitting model.

Sample 11 illustrates a NaN decision. The transmittance spectrum does not exhibit a single well-defined absorption edge, but instead shows broad shoulders and multiple changes in slope (Fig. 3c). Nevertheless, the original analysis procedure identified a locally linear region in the Tauc plot and returned $E_g$ = 4.463 eV (Fig. 3f), although this value required extrapolation by approximately 0.64 eV below the fitted Tauc region. In contrast to the ACCEPT and REPAIR_ACCEPT cases, $\ln[\alpha(h\nu)]$ below $E_g$ did not show a clear linear region but instead showed substantial curvature and changes in slope (Fig. 3i). The agent tested the permitted repairs, including relaxation of the $R^2$ criterion to 0.989 and changes in the fitting-window position, but no scientifically reliable $E_U$ was obtained. The agent's audit rationale explicitly noted the absence of a well-defined absorption edge and a consistent linear region corresponding to an Urbach tail. Concise rationales for the remaining NaN decisions, including the spectral and fitting evidence considered by the agent, are summarized in Supplementary Table S1.

## 4. Discussion

The LLM agent does not replace the numerical analysis used to extract $E_U$; instead, it supervises the deterministic procedure, evaluates the resulting spectral and fitting evidence, and decides whether to accept the original result, make a bounded adjustment, or return NaN. Importantly, this adaptation is constrained so that the measured data, the physical definition of $E_U$, and the underlying analysis model remain unchanged, while only predefined fitting parameters are adjusted when justified. This separation preserves the deterministic numerical procedure as the source of the reported physical quantity while adding an adaptive decision layer for handling cases in which fixed analysis criteria are not satisfied. In this way, the LLM agent provides scientific exception handling around the established analysis script without unnecessarily altering the physical assumptions on which the analysis is based.

The value of the adaptive agent lies not only in recovering failed analyses, but also in recognizing when a reliable value should not be reported. In the REPAIR_ACCEPT cases, limited adjustment of the fitting criteria was sufficient to obtain a stable $E_U$ estimate. By contrast, sample 11 (NaN case) illustrates why further adjustment of the fitting parameters is not always scientifically justified. The spectrum lacks a well-defined absorption edge, and the near-edge $\ln[\alpha(h\nu)]$ region does not show the linear behavior expected for an Urbach tail. Under these conditions, further relaxation of the fitting criteria or changes to the fitting window would risk producing a numerical fit without clear support from the spectral evidence. Returning NaN is thus an intentional scientific decision rather than a failure of the workflow.

In the present dataset, the cases requiring agent intervention were those in which the original analysis procedure explicitly failed to return an $E_U$ value. The study therefore does not establish whether the agent can identify cases in which an analysis completes numerically but yields a physically inappropriate result. The evaluation was also limited to one optical descriptor and one materials campaign. Future work should examine the applicability of this approach to other characterization methods and broader classes of analysis failures.

In a live SDL, the main operational benefit of such an analysis layer would be to reduce the need for synchronous human intervention when an automated analysis encounters an exception. Under a predefined protocol, ACCEPT results could proceed directly to downstream decision making, while recoverable failures could be adaptively reanalyzed and NaN could be returned for unrecoverable cases, along with an explicit rationale and audit record. Rather than requiring an expert to inspect every failed analysis before the experimental loop can continue, the system could allow exceptions to be reviewed asynchronously at an appropriate stage of the campaign. A NaN outcome could, for example, trigger a repeat measurement, alternative characterization, model-based treatment of the missing or infeasible outcome under a predefined policy,[7,48,49] or exclusion of the result from an optimization update. Thus, the role of human oversight is not eliminated but shifted from real-time intervention for individual failures toward supervision of recorded exceptions and their supporting evidence.[50] More broadly, the same approach could be extended to other characterization workflows, such as diffraction, spectroscopy, microscopy, or transport analysis, where predefined numerical routines occasionally fail and limited, auditable adaptation may allow reliable descriptor extraction.

## 5. Conclusions

We developed an LLM-agent workflow for autonomous adaptive analysis of experimental spectra and evaluated it on 203 optical transmittance spectra from a self-driving $\beta$-$Ga_2O_3$ sputter-epitaxy campaign. The original analysis procedure returned $E_U$ values for 183 spectra and failed for 20; the agent retained the 183 values returned by the baseline analysis, recovered 16 of the 20 failures through bounded adjustment of predefined fitting parameters, and returned NaN for the remaining 4 when no scientifically reliable $E_U$ could be obtained. Importantly, quantitative property extraction remained based on the deterministic analysis procedure, while the LLM agent evaluated the spectral and fitting evidence, determined whether to accept the original result, perform a bounded repair, or return NaN, and recorded an auditable rationale for each outcome. The results demonstrate a practical approach to autonomous scientific exception handling that can recover analyzable measurements without continuing to adjust the analysis merely to obtain a numerical value. By handling such exceptions adaptively and transparently, the framework could reduce the need for real-time human intervention and support more robust autonomous operation of self-driving laboratories.

## Author contributions

Y.K.W. conceived the study, designed and implemented the LLM agent, and carried out the analysis. Y.K.W. and Y.O. discussed the results. Y.K.W. wrote the manuscript with input from Y.O.

## Conflicts of interest

The authors declare no competing interests.

## Data availability

The optical transmittance spectra and original optical-analysis code from the previously reported self-driving sputtering campaign (Ref. **18**) are publicly available in the GitHub repository at https://github.com/nttcslab/interpretable-SDL. The agent instructions and a concise summary of the exception-case outcomes, together with a representative ACCEPT case, are provided in the Supplementary Information.

## Supplementary Note S1. Instructions for autonomous adaptive Urbach analysis

The following instructions were provided to the ChatGPT Workspace Agent (GPT-5.6 Sol; generation date: 25 September 2026) for the autonomous analysis of individual optical-transmission spectra. Each spectrum was analyzed in an independent fresh session. A single transmittance CSV file was attached to the session, together with the user message:
"Proceed autonomously according to your instructions with attached csv. Do not ask me for analytical guidance."

The agent was instructed to complete the analysis without requesting analytical guidance from the user.

### Workspace Agent protocol: autonomous adaptive Urbach analysis

#### Scientific objective

Treat every incoming transmission spectrum as a previously unseen experimental result from a self-driving laboratory.
Without using archived Urbach energies, publication information, future samples, historical NaN labels, or human feedback, autonomously determine whether the baseline Urbach-energy analysis should be:

- ACCEPT: the baseline result is scientifically admissible;
- REPAIR_ACCEPT: a scientifically justified bounded repair produces an admissible result; or
- NaN: no defensible Urbach energy can be obtained within the permitted repair space.

A successful numerical fit is necessary but not sufficient for scientific admissibility.

#### Scientific context available to the agent

The spectra are optical-transmission measurements of nominally 55-nm $\beta$-$Ga_2O_3$ thin films grown by sputtering on c-plane $Al_2O_3$ substrates.
The 55-nm thickness is the nominal/default thickness used by the analysis routine and should not be assumed to be an independently measured exact thickness for every individual sample.
Use this material and measurement context only to assess physical plausibility. Do not infer or search for archived results, growth-condition outcomes, or expected Eu values.
Absolute transmission magnitude may be affected by referencing, substrate/background contributions, interference, scattering, or thickness variation. Therefore:

- do not reject a spectrum solely because its absolute transmittance is low or slightly exceeds 1;
- but do consider whether the optical contrast associated with the absorption edge is sufficiently large relative to background variation or noise to support a physically meaningful Urbach analysis.

#### Inputs available to the agent

- One incoming spectrum CSV.
- Frozen baseline analysis program "Transmit_analyses_20251226.py".
- Numerical outputs, fitting diagnostics, plots, and execution errors produced by the analysis runner.
- This protocol.

The agent must not have access to:

- the publication;
- DOI or arXiv manuscript;
- campaign tables;
- archived Eu values;
- archived NaN labels;
- future spectra;
- mappings or filenames that reveal reference outcomes.

#### Required workflow for every spectrum

1. Run the frozen baseline analysis unchanged.
2. Inspect the raw spectrum together with all baseline outputs, including:

- raw transmission behavior;
    - estimated absorption edge;
    - Tauc fit;
    - candidate Urbach windows;
    - selected fitting interval, if any;
    - fit quality;
    - residual behavior;
    - point count and energy span;
    - execution errors or warnings.
3. Decide whether the returned result is scientifically admissible. A successful Python execution or high R2 is not by itself sufficient.
4. If the baseline result is defensible, return ACCEPT and the baseline Eu.
5. If the baseline result is not admissible but appears recoverable:
    - state the scientific failure diagnosis;
    - propose the smallest bounded modification needed;
    - record the exact parameter or code difference;
    - explain why the modification is scientifically justified;
    - execute the modified temporary version;
    - inspect the new fit and raw spectrum again;
    - test local stability under small neighboring fitting choices;
    - then decide REPAIR_ACCEPT or continue to another bounded repair.
6. If no defensible Eu can be obtained within the permitted modifications and retry budget, return NaN.
7. For NaN, explicitly state:
    - the NaN criterion;
    - the observed spectral feature that triggered it;
    - the repairs attempted;
    - why additional relaxation or relocation would be scientifically unjustified.
8. Save a structured audit record for the spectrum.

**Core scientific admissibility requirements**
Assess the following jointly. Do not use any single numerical threshold as the sole justification.

**A. Localized absorption edge**
A physically recognizable and localized high-energy absorption-edge feature must be present.
A broad drift, multiple unrelated shoulders, isolated curvature, or a spectrum without a distinguishable edge should not be converted into a finite Eu merely because a linear interval can be found somewhere.

**B. Edge-connectedness of the Urbach region**
The candidate Urbach region must be physically connected to the pre-edge region immediately below the estimated absorption edge.
A repaired fitting window must not be relocated to a lower-energy, spectrally disconnected region solely because that region gives:

- higher R2;
- better monotonicity;
- smaller residuals; or
- better local numerical stability.

Local linearity and local stability are necessary but not sufficient.
The candidate must remain interpretable as the exponential tail associated with the same absorption edge used by the analysis.

**C. Monotonicity and exponential-tail behavior**
The selected region should show a physically coherent positive-slope trend in ln(alpha) versus photon energy.
Point-scale noise is acceptable when the broader trend is coherent, but a fit should not be accepted when it relies on oscillations, isolated short fragments, or curvature inconsistent with an exponential tail.

**D. Fit extent and support**

Consider:

- number of fitting points;
- energy span;
- R2;
- residual magnitude and structure;
- positive-step fraction;
- relationship to the neighboring raw-spectrum shape.

Do not accept a short high-R2 interval automatically if a broader neighboring region shows strong curvature or contradicts the interpretation.

**E. Local stability**

Test small neighboring choices of fitting boundaries or widths.

A defensible result should not change drastically under small nearby choices.

However, stability of Eu within an arbitrarily relocated spectral region does not establish physical validity. Edge-connectedness must be satisfied first.

**F. Optical signal strength and physical plausibility**

Consider whether the absorption-edge signal is sufficiently distinguishable from background variation, noise, or weak optical contrast.

For nominally 55-nm β-Ga2O3/c-plane Al2O3 spectra, a very small transmission change may make an apparently precise Eu physically less reliable if the inferred tail is supported by only a weak optical signal.

Do not impose a fixed absolute-transmittance cutoff.

Instead assess whether:

- the edge-associated change is visibly resolvable;
- the fitted change is large compared with local fluctuations/background;
- the inferred tail remains credible when viewed in the raw spectrum.

If signal amplitude is questionable but the fit is otherwise admissible, record this explicitly in the audit rationale.

**Safety and reproducibility constraints**

- Never modify the raw spectrum.
- Never overwrite the frozen baseline source file.
- Apply changes only to a temporary copy or via explicit parameter overrides.
- Prefer existing exposed fitting parameters over source-code edits.
- Do not tune toward any expected or reference Eu value.
- Do not search the web or external applications.
- Maximum repair attempts per spectrum: 3.
- Each new spectrum starts from exactly the same frozen baseline version.
- Modifications from one sample must never propagate to another sample.
- Do not use previous samples' decisions to determine the current sample's result.

**Allowed first-line repair space**

Repairs must remain small and scientifically interpretable.

**Urbach R2 threshold**

The baseline value is 0.995.

A limited relaxation may be tested when the raw spectrum and a broad edge-connected candidate tail are otherwise scientifically convincing.

Do not lower the Urbach R2 threshold below 0.989 in this benchmark.

A lower R2 must never be used simply to maximize the number of finite outputs.

**Offset from the estimated absorption edge**

The baseline end offset is 0.03 eV.

First-line repair may vary the Urbach end offset only within:

0.00 eV <= end_offset <= 0.08 eV.
Do not move the Urbach endpoint farther than 0.08 eV below the estimated edge to rescue a failed fit.
If no defensible edge-connected Urbach region exists within this range, prefer NaN rather than progressively searching farther from the absorption edge.

**Fitting-window extent**
Small changes to the initial window width or use of the existing baseline expansion logic are permitted, provided that:

- the selected region remains edge-connected;
- the region is not chosen merely because it maximizes R2;
- the fit remains physically interpretable as an Urbach tail.

The existing maximum expansion of 0.40 eV should not be exceeded.

**Multiple simultaneous changes**
Prefer changing one exposed parameter at a time.
Do not simultaneously relax multiple criteria unless the preceding attempt establishes why a single-parameter repair is insufficient.
Every additional change must have a separate scientific rationale.

**Source-code modification**
A source-code patch should be exceptional.
Before modifying source logic, state:

- why the existing exposed parameters are insufficient;
- why the proposed modification preserves the physical definition of Eu;
- why the change is not merely a way to force a numerical fit.

The raw measurements, Urbach relation, property definition, and frozen original source remain immutable.

**NaN principle**
NaN is a valid scientific outcome.
Return NaN when the requested Eu is not supportable under the permitted analysis, including situations such as:

- no localized absorption edge;
- no edge-connected monotonic exponential-tail region;
- candidate regions dominated by curvature or oscillation;
- apparently good fits that occur only in spectrally disconnected regions;
- a result requiring movement beyond the allowed edge-offset range;
- a result requiring excessive threshold relaxation;
- substantial instability under nearby admissible fitting choices;
- optical contrast too weak to distinguish the candidate tail reliably from background/noise;
- further changes would amount to searching for any linear segment rather than testing the physical Urbach model.

Do not interpret NaN as proof of failed film growth, absence of material, or absence of an absorption edge.
It means only that a defensible Urbach energy cannot be assigned under this analysis protocol.
Infrastructure or execution failure must be recorded separately and must not be converted into a scientific NaN.

**Required audit output**
For each spectrum save:

- anonymized sample ID;
- spectrum hash;
- frozen baseline source hash;
- baseline execution status;
- baseline Eu, if any;
- estimated Eg;
- raw-spectrum summary;
- baseline fitting diagnostics;

- candidate-window diagnostics;
- assessment of localized absorption edge;
- assessment of edge-connectedness;
- assessment of optical signal strength / contrast;
- agent diagnosis;
- decision after each attempt;
- every exact parameter change or source-code diff;
- scientific rationale for every change;
- rerun outputs;
- final fitting diagnostics;
- local stability results;
- final decision: ACCEPT, REPAIR_ACCEPT, or NaN;
- final Eu when applicable;
- explicit NaN criterion and rationale when applicable;
- number of repair attempts.

For every ACCEPT or REPAIR_ACCEPT, the final rationale must explicitly state why:

1. the candidate is connected to the absorption edge;
2. the fit is physically consistent with an Urbach tail;
3. local stability is adequate;
4. the result is not merely a high-R2 fit to an arbitrary linear fragment.

For every NaN, explicitly state why further repair would become scientifically arbitrary.

**Batch ZIP behavior**

When a ZIP archive containing multiple spectra is uploaded:

1. Extract eligible spectrum CSV files.
2. Process each spectrum independently using this complete protocol.
3. Reset to the identical frozen baseline before every spectrum.
4. Do not propagate repairs, criteria, or inferred parameter preferences between samples.
5. Produce one audit JSON per spectrum.
6. Produce diagnostic figures for exception cases when possible.
7. Produce a batch summary spreadsheet containing at minimum:
    - source file;
    - anonymized sample ID;
    - Eg;
    - Eu;
    - final decision;
    - repair attempts;
    - exact modification;
    - final R2;
    - fitting-point count;
    - fitting range;
    - concise scientific rationale;
    - NaN criterion when applicable.

Do not stop for confirmation between spectra.

**Automatic execution behavior**

When the user uploads either:

- one eligible spectrum CSV; or
- a ZIP archive containing eligible spectrum CSV files,

begin the autonomous Urbach-analysis workflow immediately.
Do not ask the user which action to perform.
Do not ask for scientific guidance during processing.
Complete the ACCEPT / REPAIR_ACCEPT / NaN decision autonomously according to this frozen protocol.

**Benchmark rule**

Do not inspect archived Eu values, historical NaN labels, publication results, campaign tables, or any reference mapping until all benchmark decisions have been completed and frozen.

Do not alter this protocol based on the results of individual benchmark samples once the full rerun has begun.

**Supplementary Table S1. Autonomous analysis outcomes and concise audit rationales.**

The table lists the 20 spectra for which the original analysis procedure did not return an Urbach-energy value, together with sample 24 as a representative ACCEPT case.

| Sample | Final decision | $E_U$ (meV) | Parameter adjustment | Concise audit rationale |
|---|---|---|---|---|
| 24 | **ACCEPT** | 183.0 | None | The original result was retained. The fitted region lay immediately below the same absorption edge used for $E_g$, $\ln[\alpha(h\nu)]$ was monotonic with $R^2$ = 0.99713, the transmission change was clearly larger than local measurement noise, and nearby boundary choices gave $E_U$ = 172.6–192.7 meV. |
| 11 | **NaN** | NaN | Unsuccessful, 3 attempts: Urbach $R^2$ threshold: 0.995 → 0.989; high-energy boundary: $E_g$ - 0.03 → $E_g$; high-energy boundary: $E_g$ - 0.03 → $E_g$ - 0.08 eV | The spectrum lacked a localized absorption edge; the Tauc $E_g$ required a long extrapolation, and all allowed near-edge $\ln[\alpha(h\nu)]$ candidates were curved or nonqualifying. Further threshold relaxation, endpoint movement, or relocation of the fitting region was not scientifically justified. |
| 33 | **REPAIR_ACCEPT** | 259.8 | Urbach $R^2$ threshold: 0.995 → 0.989 | A broad candidate immediately below the absorption edge became admissible after a bounded $R^2$ relaxation. The accepted interval contained 33 points over 0.3165 eV with $R^2$ = 0.98904; nearby boundary perturbations gave $E_U$ = 251.1–273.0 meV, within about 5.1% of the selected value. Mild curvature was retained as a caveat. |
| 75 | **REPAIR_ACCEPT** | 519.3 | High-energy boundary: $E_g$ - 0.03 → $E_g$ - 0.01 eV | Moving the upper-energy boundary closer to the same absorption edge recovered an 8-point monotonic fit with $R^2$ = 0.99718 without relaxing the fit-quality criterion. The transmission change was clearly larger than local measurement noise, and nearby windows gave $E_U$ = 519.3–558.0 meV. |
| 85 | **REPAIR_ACCEPT** | 464.9 | Fitting-window width: 0.08 → 0.06 eV | The default window-search increment skipped an intermediate-width, edge-connected region. Refining the search increment recovered a 13-point monotonic fit with $R^2$ = 0.99543 while retaining the original $R^2$ criterion; nearby qualifying fits gave $E_U$ = 457.0–464.9 meV. |
| 89 | **REPAIR_ACCEPT** | 561.2 | Urbach $R^2$ threshold: 0.995 → 0.993 | A small $R^2$ relaxation admitted a broad 27-point, 0.237 eV region directly below the edge. The positive $\ln[\alpha(h\nu)]$ trend, resolved transmission change, and neighboring boundary tests supported the result; the maximum neighboring $E_U$ deviation was about 6.1%. |
| 155 | **NaN** | NaN | Unsuccessful, 2 attempts: Tauc $R^2$ threshold: 0.995 → 0.994; then Urbach $R^2$ threshold: 0.995 → 0.989 | A small Tauc-threshold relaxation recovered a provisional $E_g$, but no allowed near-edge interval supported an Urbach tail. The strongest allowed candidate reached only $R^2$ = 0.93891, had systematic residual structure, and only 62.5% positive steps. Further relaxation or relocation would exceed the permitted repair space. |
| 157 | **NaN** | NaN | Unsuccessful, 3 attempts: Urbach $R^2$ threshold: 0.995 → 0.989; high-energy boundary: $E_g$ - 0.03 → $E_g$; then both changes combined | A real absorption edge and an acceptable Tauc estimate were present, but $\ln[\alpha(h\nu)]$ immediately below the edge remained curved. The strongest 8-point candidate reached only $R^2$ = 0.98096, below the 0.989 repair floor, and adding neighboring support reduced linearity and changed the inferred slope. |
| 158 | **REPAIR_ACCEPT** | 424.2 | Urbach $R^2$ threshold: 0.995 → 0.994 | A 0.001 $R^2$ relaxation admitted the same default-endpoint 26-point, 0.230 eV candidate, with $R^2$ = 0.99420. The fitted region showed a coherent positive slope, strong raw optical contrast, small residuals, and approximately 1% $E_U$ variation under nearby endpoint and boundary changes. |
| 166 | **REPAIR_ACCEPT** | 159.4 | Urbach $R^2$ threshold: 0.995 → 0.989 | The unchanged baseline window became admissible after threshold relaxation. The 8-point, 0.069 eV interval had $R^2$ = 0.99427 and every $\ln[\alpha(h\nu)]$ step was positive; nearby admissible windows kept $E_U$ within 7.1% of the selected value without relocating the fit. |
| 168 | **REPAIR_ACCEPT** | 233.9 | Urbach $R^2$ threshold: 0.995 → 0.994 | A minimal $R^2$ relaxation admitted a 17-point edge-anchored fit with $R^2$ = 0.99489. The sampled $\ln[\alpha(h\nu)]$ trend was strictly monotonic, the raw contrast was well resolved, and neighboring offsets, widths, and boundary choices supported local stability. |

| Sample | Final decision | $E_U$ (meV) | Parameter adjustment | Concise audit rationale |
|---|---|---|---|---|
| 170 | **REPAIR_ACCEPT** | 184.8 | Urbach $R^2$ threshold: 0.995 → 0.989 | Relaxing only the $R^2$ criterion to the protocol floor admitted a broad 25-point, 0.226 eV edge-connected interval with $R^2$ = 0.98999. The fit had a coherent positive slope and strong optical contrast; nearby sampled boundaries and neighboring broad-window endpoints kept $E_U$ clustered near 0.185 eV. |
| 171 | **REPAIR_ACCEPT** | 484.9 | 2 attempts: high-energy boundary: $E_g$ - 0.03 → $E_g$; then Urbach $R^2$ threshold: 0.995 → 0.989 | Moving the endpoint to $E_g$ alone did not recover a fit; subsequent relaxation to the $R^2$ floor admitted a broad 35-point, 0.312 eV edge-connected interval with $R^2$ = 0.98935. Admissible local perturbations gave $E_U$ = 484.9–497.7 meV with a coefficient of variation of 0.73%. |
| 175 | **REPAIR_ACCEPT** | 277.6 | Urbach $R^2$ threshold: 0.995 → 0.994 | A minimal 0.001 $R^2$ relaxation admitted a broad 35-point, 0.312 eV interval at the unchanged edge offset. The fit showed a positive ln[$\alpha(h\nu)$] slope and small residuals, and neighboring offsets and widths changed $E_U$ by less than about 1% in the primary stability tests. |
| 177 | **REPAIR_ACCEPT** | 1573.2 | 3 attempts; final: high-energy boundary: $E_g$ - 0.03 → $E_g$ plus Urbach $R^2$ threshold: 0.995 → 0.989 | Only the combined endpoint and threshold repair produced a finite result: a 9-point, 0.072 eV fit with $R^2$ = 0.99072 directly below $E_g$. Nearby $E_U$ values changed by roughly 6% or less, but the support was local and broader windows showed curvature; this limitation was retained as an audit caveat. |
| 182 | **REPAIR_ACCEPT** | 195.6 | Urbach $R^2$ threshold: 0.995 → 0.989 | Threshold relaxation admitted the immediate pre-edge 0.16 eV interval. The fitted region showed a coherent positive slope with 93.75% positive steps and resolved raw optical contrast; two-point boundary perturbations gave $E_U$ = 189.2–203.8 meV with about 2.2% relative dispersion. |
| 184 | **REPAIR_ACCEPT** | 182.0 | 3 attempts; final: high-energy boundary: $E_g$ - 0.03 → $E_g$ - 0.06 eV plus Urbach $R^2$ threshold: 0.995 → 0.989 | Neither the threshold relaxation nor the endpoint change succeeded alone. Their bounded combination recovered a 26-point, 0.230 eV edge-connected fit with $R^2$ = 0.98955; independent boundary shifts gave $E_U$ = 176.0–187.1 meV, within 3.3% of the selected value. |
| 186 | **NaN** | NaN | Unsuccessful, 3 attempts: Urbach $R^2$ threshold: 0.995 → 0.989; high-energy boundary: $E_g$ - 0.03 → $E_g$; then both changes combined | A localized and optically resolvable edge was present, but the connected pre-edge ln[$\alpha(h\nu)$] region remained curved or oscillatory. The best allowed fit reached only $R^2$ = 0.97808, and all bounded repairs failed. A finite result would require exceeding the $R^2$ or endpoint limits or selecting a disconnected fragment. |
| 188 | **REPAIR_ACCEPT** | 208.1 | High-energy boundary: $E_g$ - 0.03 → $E_g$ - 0.04 eV | A 0.01 eV endpoint shift recovered a 17-point, 0.149 eV edge-connected fit with $R^2$ = 0.99562 without relaxing the $R^2$ criterion. ln[$\alpha(h\nu)$] was monotonic, and nearby boundaries, offsets, and widths changed $E_U$ by no more than about 6.0%. |
| 190 | **REPAIR_ACCEPT** | 146.1 | Urbach $R^2$ threshold: 0.995 → 0.989 | Threshold relaxation admitted a 16-point, 0.1445 eV fit with $R^2$ = 0.99114 on the broad monotonic edge. Seven of nine nearby windows also met the repaired $R^2$ criterion and gave $E_U$ = 142.3–149.0 meV, with a maximum deviation of 2.6% from the selected value. |
| 197 | **REPAIR_ACCEPT** | 585.4 | 3 attempts; final: high-energy boundary: $E_g$ - 0.03 → $E_g$ plus Urbach $R^2$ threshold: 0.995 → 0.989 | The threshold and endpoint changes each failed alone, but their combination admitted the full 0.40 eV edge-adjacent interval: 44 points with $R^2$ = 0.98922. Nearby boundary choices gave $E_U$ = 582.6–601.2 meV, a 1.15% coefficient of variation and 2.69% maximum deviation. |